\documentclass[letterpaper,aps,pra,showpacs,floats,nofootinbib,superscriptaddress,reprint]{revtex4-1}
\usepackage[latin1]{inputenc}
\usepackage[T1]{fontenc}
\usepackage{amsmath,amssymb,mathbbol,dsfont}

\usepackage{graphicx,color}
\usepackage{hyperref}

\usepackage{placeins}
\usepackage{microtype}

\newcommand{\titlestr}{Time delays for atto\-second streaking in photoionization of neon}

\hypersetup{colorlinks,hypertexnames=true,pdftitle={\titlestr},linkcolor=blue,citecolor=blue,urlcolor=blue}

\renewcommand{\Re}{\operatorname{Re}}

\newcommand{\Schro}{Schr\"o\-din\-ger }
\newcommand{\etal}{\emph{et~al.\ }}
\newcommand{\ie}{i.e., }

\newcommand{\abs}[1]{\left|#1\right|}
\newcommand{\bra}[1]{\langle#1|}
\newcommand{\ket}[1]{|#1\rangle}
\newcommand{\braket}[2]{\langle#1|#2\rangle}
\newcommand{\level}[3]{{}^{#1}\!{#2}^{\textrm{#3}}}

\newcommand{\op}[1]{\mathrm{\hat{#1}}}
\newcommand{\subscr}[1]{_{\scriptstyle\textrm{#1}}}
\newcommand{\ddE}{\frac{\partial}{\partial E}}
\newcommand{\hw}{\hbar\omega}

\newcommand{\tEWS}{t\subscr{EWS}}
\newcommand{\tCLC}{t\subscr{CLC}}
\newcommand{\tdLC}{t\subscr{dLC}}
\newcommand{\tS}  {t\subscr{S}}
\newcommand{\tEWSC}{\tEWS^{\scriptstyle\textrm{C}}}
\newcommand{\tEWSS}{\tEWS^{\scriptstyle\textrm{S}}}
\newcommand{\Eph}{E\subscr{ph}}

\newcommand{\eV}{\,\text{eV}}
\newcommand{\au}{\,\text{a.u.}}
\newcommand{\Wcm}{\,\text{W}/\text{cm}^2}
\newcommand{\as}{\,\text{as}}
\newcommand{\fs}{\,\text{fs}}
\newcommand{\Hep}{\ensuremath{\text{He}^{+}}}

\begin{document}
\title{\titlestr}
\author{Johannes Feist}
\email{johannes.feist@uam.es}
\affiliation{ITAMP, Harvard-Smithsonian Center for Astrophysics, Cambridge, Massachusetts 02138, USA}
\affiliation{Departamento de F\'isica Te\'orica de la Materia Condensada, Universidad Aut\'onoma de Madrid, 28049 Madrid, Spain, EU}

\author{Oleg Zatsarinny}
\affiliation{Department of Physics and Astronomy, Drake University, Des Moines, Iowa 50311, USA}

\author{Stefan Nagele}
\email{stefan.nagele@tuwien.ac.at}
\affiliation{Institute for Theoretical Physics, Vienna University of Technology, 1040 Vienna, Austria, EU}

\author{Renate Pazourek}
\affiliation{Institute for Theoretical Physics, Vienna University of Technology, 1040 Vienna, Austria, EU}

\author{Joachim Burgd\"orfer}
\affiliation{Institute for Theoretical Physics, Vienna University of Technology, 1040 Vienna, Austria, EU}

\author{Xiaoxu Guan}
\affiliation{Department of Physics and Astronomy, Drake University, Des Moines, Iowa 50311, USA}

\author{Klaus Bartschat}
\affiliation{ITAMP, Harvard-Smithsonian Center for Astrophysics, Cambridge, Massachusetts 02138, USA}
\affiliation{Department of Physics and Astronomy, Drake University, Des Moines, Iowa 50311, USA}

\author{Barry I. Schneider}
\affiliation{Office of Cyberinfrastructure, National Science Foundation, Arlington, Virginia 22230, USA}

\date{\today}


\begin{abstract}
We revisit the time-resolved photoemission in neon atoms as probed by atto\-second streaking. 
We calculate streaking time shifts for the emission of $2p$ and $2s$ electrons and compare the relative delay as measured in a 
recent experiment by Schultze \etal [Science 328, 1658 (2010)].
The \hbox{$B$-spline} \hbox{$R$-matrix} method is employed to calculate accurate Eisenbud-Wigner-Smith 
time delays from multi-electron dipole transition matrix elements for photoionization.
The additional laser field-induced time shifts in the exit channel are obtained from separate, 
time-dependent simulations of a full streaking process by solving the time-dependent \Schro equation 
on the single-active-electron level. 
The resulting accurate total relative streaking time shifts between $2s$ and $2p$ emission lie well below the experimental data. 
We identify the presence of unresolved shake-up satellites in the experiment as
a potential source of error in the determination of streaking time shifts.
\end{abstract}

\maketitle

\section{Introduction}\label{sec:intro}
The photoelectric effect, \ie the emission of an electron after the absorption of a photon, 
is one of the most fundamental processes in the interaction of light with matter.
Progress in the creation of ultrashort light pulses during the past decade \cite{Corkum07,KraIva2009,ChaCor2010} 
has enabled the time-resolved study of photoemission with atto\-second ($1\as = 10^{-18}\,$s) precision.
In a pioneering experimental work, Schultze \etal~\cite{SchFieKar2010} reported a time delay of $21 \pm 5\as$ 
between the emission of $2s$ and $2p$ electrons from neon, measured using the atto\-second streaking 
technique \cite{HenKieSpi2001, DreHenKie2001, ItaQueYud2002, KieGouUib2004}.
However, the measured relative delay has not yet been quantitatively confirmed by theory, even 
though several time-dependent as well as time-independent state-of-the-art methods have already been 
applied to the problem \cite{SchFieKar2010,KheIva2010,MooLysPar2011,NagPazFei2012,Khe2013}.  

Previous time-dependent studies have aimed at a simulation of the streaking spectrogram
 \cite{SchFieKar2010,MooLysPar2011,NagPazFei2012}, whereas the time-independent 
approaches \cite{SchFieKar2010,KheIva2010,Khe2013} have focused on accurate calculations of the 
quantum-mechanical Eisenbud-Wigner-Smith (EWS) delay \cite{Eis1948,Wig1955,Smi1960} from the 
dipole-matrix elements for the photoionization process, \ie the group delay of the 
photoelectron wavepacket~\cite{deCNus2002}.
The latter methods allow for an accurate description of electronic correlations in the 
photo\-ionization process, but they ignore the influence of the infrared (IR) field on the extracted time shifts. 
For the time-dependent simulations the situation is reversed. 
While they account for the influence of the IR streaking field on the photoemission process, 
their inclusion of electron-electron correlation is incomplete.
So far only simulations for one and two active electrons in model systems \cite{NagPazFei2011,NagPazFei2012} 
and time-dependent \hbox{$R$-matrix} calculations for Ne with restricted basis sizes \cite{MooLysPar2011} have become available.

The starting point of the present investigation is the key observation 
\cite{ZhaThu2010,NagPazFei2011,ZhaThu2011,PazFeiNag2012,NagPazFei2012,PazNagBur2013} 
that the contributions to the total streaking time delay $\tS$, due to the intrinsic 
atomic EWS delay and to the IR streaking field, are strictly additive with sub-attosecond precision. 
Therefore, both contributions can be determined independently of each other in separate treatments, both featuring high precision. 

In this contribution, we implement such an approach for calculating the total streaking time shifts $\tS$ for the
neon atom by using the \hbox{$B$-spline} \hbox{$R$-matrix} (BSR) method \cite{Zat2006,ZatFro2009} for the EWS delays and accurate 
time-dependent \emph{ab initio} one- and two-active electron simulations \cite{TonChu1997,NagPazFei2011,PazFeiNag2012} for 
simulating IR-field-induced time shifts containing a Coulomb-laser, $\tCLC$ \cite{ZhaThu2010,NagPazFei2011,PazNagBur2013}, 
and a dipole-laser coupling contribution, $\tdLC$ \cite{BagMad2010,*BagMad2010Err,PazFeiNag2012,PazNagBur2013}. 
This procedure has the advantage that the calculation of both contributing parts can be independently optimized. 
We find the resulting time delay, $\Delta \tS = \tS^{\scriptstyle{(2p)}}-\tS^{\scriptstyle{(2s)}}$, to be about a 
factor of 2 smaller than the experiment, which seems well outside the theoretical uncertainty of our calculation. 
We furthermore explore the possible influence of unresolved shake-up channels in the experiment as a potential 
source of error in the determination of $\Delta \tS$. 

This paper is organized as follows. In \autoref{sec:model} we describe our method. 
This is followed by a presentation and discussion of our results for $\tEWS$, $\tCLC$, and the total streaking 
time delay $\Delta \tS$ in \autoref{sec:results}. 
Possible corrections due to contamination by shake-up channels are discussed in \autoref{sec:results_shake}, 
followed by a brief summary (\autoref{sec:summary}).
Atomic units are used throughout unless explicitly stated otherwise.

\section{Theoretical Approach}\label{sec:model}
Time-resolved atomic photoionization in an attosecond-streaking setting involves two light fields,
namely the ionizing isolated attosecond pulse in the extreme ultraviolet (XUV) range of the spectrum,
$\vec F\subscr{XUV}(t)$, and the streaking (or probing) IR field $\vec F\subscr{IR}(t)$,
\begin{equation}\label{eq:fields}
\vec F(t)= \vec F\subscr{XUV}(t) + \vec F\subscr{IR}(t)\, . 
\end{equation}
By varying the temporal overlap between $\vec F\subscr{IR}$ and $\vec F\subscr{XUV}$, timing information on 
the attosecond scale can be retrieved \cite{SchFieKar2010,CavMueUph2007}.
While the XUV field is weak and can be safely treated within first-order perturbation theory, the streaking 
field is moderately strong, such that the continuum state of the liberated electron is strongly perturbed,
while the initial bound state is not
yet appreciably ionized by $\vec F\subscr{IR}$. 
This gives rise to the characteristic streaking spectrogram (see below), with a time-dependent momentum shift of
the free electron proportional to the time-shifted vector potential of the IR field, i.e., $\Delta \vec p(t)\propto \vec A\subscr{IR}(t+\tS)$.
Here, $t$ is the peak time of the attosecond pulse, while $\tS$ is the \emph{absolute} streaking time shift in time-resolved photoionization.
To emphasize that $\tS>0$ corresponds to \emph{delayed} emission and due to its relation to the EWS delay,
streaking time shifts are often also called streaking \emph{delays}. Both notations will be used interchangeably in the following.
Note that in experiment, the \emph{relative} streaking time shift or delay $\Delta \tS$ between two different ionization channels is measured.

The total absolute streaking delay $\tS$ can be decomposed
with sub-attosecond precision into a contribution from the intrinsic EWS time delay $\tEWS$ for ionization 
by the XUV pulse in the absence of a probing field and contributions that stem from the combined interaction 
of the electron with the streaking IR field and the long-range fields of the residual ion, $\tCLC$ and $\tdLC$. Specifically,
\begin{equation}\label{eq:delay_rel_all}
 \tS = \tEWS + \tCLC + \tdLC \, .
\end{equation}
The Coulomb-laser coupling (CLC) time shift $\tCLC$ results from the interplay between the streaking and Coulomb fields.
It is universal in the sense that it depends only on the frequency $\omega\subscr{IR}$ of the streaking field, 
the strength of the Coulomb field ($Z\!=\!1$ for single ionization), and the final energy of the emitted electron, 
but is independent of the strength of the IR field and of short-range admixtures to the atomic potential. 
It can be determined with sub-attosecond precision by the numerical solution of the time-dependent \Schro equation 
at the single-active electron level. 
Alternatively, it can be approximately determined from classical trajectory simulations \cite{NagPazFei2011,PazNagBur2013} 
or the eikonal approximation \cite{ZhaThu2010}. 
Closely related, a similar time shift $\tau\subscr{cc}$ describing lowest-order continuum-continuum coupling appears in the complementary 
interferometric ``RABBIT'' technique \cite{ZhaThu2010,DahCarLin2012,DahHuiMaq2012,DahGueKlu2012,CarDahArg2013}. 

In the presence of near-degenerate initial or final states with non-zero dipole moments (\ie linear Stark shifts), 
an additional IR-field-induced time shift, the dipole-laser coupling (dLC) contribution $\tdLC$ appears (\autoref{eq:delay_rel_all}) \cite{BagMad2010,*BagMad2010Err,PazFeiNag2012,PazNagBur2013}.
The latter is also independent of short-range interactions, but depends on the strength of the dipole moment of the 
initial atomic or final ionic state, the IR frequency $\omega\subscr{IR}$, and the final energy of the emitted electron. 
For non-hydrogenic systems, it additionally depends on the residual splitting $\Delta E$ of the dipole-coupled near-degenerate states. 

A promising strategy for obtaining precise theoretical predictions for total streaking time shifts is thus to 
combine time-independent state-of-the-art calculations of atomic dipole matrix elements for many-electron systems 
governing $\tEWS$ with TDSE solutions on the one- and two-active electron level to accurately determine $\tCLC$ and $\tdLC$. 
By comparison, it is still extremely challenging to obtain converged solutions of the time-dependent \Schro{} equation 
for many-electron atoms in moderately strong IR fields. 

The EWS time delay is given by the energy derivative of the dipole transition matrix 
element between the initial bound state $\ket{\Psi_i}$ and the final continuum state $\ket{\Psi_f}$ of the many-electron system,
\begin{equation}
 \label{eq:EWSdelay}
 \tEWS(E,\Omega) = \ddE \arg\left[\bra{\Psi_f(E,\Omega)} \op z \ket{\Psi_i}\right] \, ,
\end{equation}
where $E$ is the energy of the photo\-electron emitted in the direction $\Omega=(\theta,\varphi)$ and 
$\op z$ is the electric dipole operator for linear polarization.
For Ne the initial state $\Psi_i$ is given by the $1s^2 2s^2 2p^6$ electronic configuration while the final 
state in the experiment \cite{SchFieKar2010} is assumed to consist of a free continuum electron and either a $1s^2 2s^22p^5$ 
ionic state (approximately corresponding to ionization of a $2p$ electron) or a $1s^2 2s2p^6$ ionic state (approximately corresponding to ionization of a $2s$ electron). 
As the core remains unaffected, for brevity we will omit the $1s^2$ electrons in the state labels below.
In the experiment, the emitted electrons were collected along the laser polarization axis, 
\ie the $z$-axis. 
We thus calculate the EWS time delay according to \autoref{eq:EWSdelay} for $\theta = 0$ or~$\pi$. 
Due to the cylindrical symmetry of the system, the results are independent of $\varphi$.

Employing $LS$-coupling, the transition matrix element between the initial state of symmetry $\level{1}{S}{e}$ and a final state
with symmetry $\level{1}{P}{o}$ is given by
\begin{multline}
\label{eq:matel}
\bra{\Psi_{\alpha}(E,\Omega)} \op z\ket{\Psi_i} = 
\sum_{\ell,m} e^{i\left[\sigma_\ell(E)-\ell\pi/2+\delta_\ell(E)\right]} \\
\abs{\bra{\Psi_{\alpha}(E,\ell)}\op z\ket{\Psi_i}} \braket{L\,\text{-}m;\ell m}{10}  Y^\ell_m(\Omega), 
\end{multline}
where $\alpha$ is the label of the final ionic state with total angular momentum~$L$,
$Y^\ell_m(\Omega)$ is a spherical harmonic, and $\braket{L\,\text{-}m;\ell m}{10}$ denotes a standard Clebsch-Gordan coefficient.

The sums over $\ell$ and $m$ include all allowed angular momenta and their $z$-projections of the free electron. 
We have explicitly separated the modulus and phase of the transition matrix element for each $\ell$ of the continuum state. 
The phase can be decomposed into the long-range 
Coulomb phase $\sigma_\ell(E)=\arg \Gamma(\ell+1+i\eta)$ with $\eta = -1/\sqrt{2E}$, the phase due to the centrifugal 
potential, $-\ell\pi/2$, and the phase shift $\delta_\ell(E)$ containing the effects of short-range interactions due to electron correlations. 
For emission along the laser polarization axis, we have
$Y^\ell_m(\Omega) \to \delta_{m,0}\sqrt{(2\ell+1)/(4\pi)}$, and \autoref{eq:matel} simplifies to
\begin{multline}
\label{eq:matel_z}
\bra{\Psi_{\alpha}(E,\theta\!=\!0)} \op z\ket{\Psi_i} = 
\sum_\ell e^{i\left[\sigma_\ell(E)-\ell\pi/2+\delta_\ell(E)\right]} \\
\sqrt{\frac{2\ell+1}{4\pi}} \abs{\bra{\Psi_{\alpha}(E,\ell)}\op z\ket{\Psi_i}} \braket{L0;\ell0}{10} \,.
\end{multline}

The complex-valued dipole matrix elements $\bra{\Psi_{\alpha}(E,\ell)}\op z\ket{\Psi_i}$ are calculated using the
BSR-PHOT program~\cite{Zat2006}. 
It utilizes
the BSR method with expansions based on multi-configuration Hartree-Fock (MCHF) states with non\-orthogonal sets of 
one-electron orbitals \cite{Zat2006,ZatFro2009}. 
In the absence of shake-up, the three exit channels of interest for ionization out of the $2s$ or the $2p$ subshell 
are $\ket{2s^22p^5 Es},\ket{2s^22p^5 Ed}$, and $\ket{2s2p^6 Ep}$ (all with symmetry $\level1Po$).  
In the current set of calculations these channels are represented by 
MCHF states for the ionic parts $\ket{2s^22p^5}$ and $\ket{2s2p^6}$, multiplied by
a \hbox{$B$-spline} basis expansion for the free electron. This model should provide most, if not all, of the physically 
relevant effects.  However, in order to evaluate the quality of the simple three-channel BSR model for the $\tEWS$, 
we performed two more extensive calculations.  The first model included all $n\!=\!3$ ionic excitations,
which amounts to 36 additional ionic target states with configurations $2s^22p^43s$, $2s^22p^43p$, $2s^22p^43d$,
$2s2p^53s$, $2s2p^53p$, $2s2p^53d$, $2s^02p^63s$, $2s^02p^63p$, and $2s^02p^63d$ (see \autoref{tab:states}),
and results in up to 57 coupled channels. The second model included pseudo\-states to
account for the polarization of the residual ion by the outgoing photo\-electron. 
We note that the model with 38 target states also accurately represents the series of $2s^2$-hole doubly excited states 
that have been previously discussed by Komninos \etal~\cite{KomMerNic2011}. 
These resonances are long-lived ($>\!100\fs$) and thus narrow, and have small dipole transition probabilities from 
the ground state. Hence they are not resolved in typical photoelectron spectra.
If they were excited with significant probabilities, they would appear in streaking spectrograms as sidebands 
with delay-time dependent modulations~\cite{DreHenKie2002}. 
The absence of these sidebands in the experimental streaking spectra \cite{SchFieKar2010} indicates that these 
resonances do not efficiently contribute to the observed time delay. 
We therefore 
remove them by smoothing the phase of the dipole matrix elements before calculating its derivative. 
To this end, we fit the phase $\delta_\ell(E)$ to a $4$th-order polynomial in the energy range of interest.
The total EWS delay is given by 
\begin{align}\label{eq:tews_decomposition}
\tEWS &= \ddE\left[\sigma_\ell(E)+\delta_\ell(E)\right]\nonumber\\ 
 & = \tEWSC(E)+\tEWSS(E)\, ,
\end{align}
consisting of the Coulomb EWS delay, $\tEWSC(E)$, and the delay due to the short-range contributions, $\tEWSS(E)$. 
In turn, the total streaking time shifts are determined by adding the IR-field-induced corrections (\autoref{eq:delay_rel_all}).

\squeezetable{
\begin{table}[tb]
\vspace{-0.23cm}
\caption{Ionic target states included in the 38-target calculations, along with their energy $E$
relative to the ionization potential $I_p$, ionization probability $P\subscr{ion}$,
expected streaking time delays $\tS=\tCLC+\tEWS$, and additional contribution $\tdLC$
(only shown when non-zero). All values are evaluated for electron emission along the 
laser polarization axis and for a photon energy $\Eph=106\eV$. States with $P\subscr{ion}=0$
do not contribute either due to symmetry or because their ionization threshold is
above the photon energy (last four states).}\label{tab:states}
\begin{ruledtabular}
\begin{tabular}{lrrrr}
 state & $E$ [eV] & $P\subscr{ion}$ [arb.u.] & $\tS$ [as] & $\tdLC$ [as] \\
\hline
$2p$ [$2p^5(^2P^o)$]   &  0.000 &    4.90927 &   -3.194 &        \\
$2s$ [$2s2p^6(^2S^e)$] & 26.795 &    0.64966 &  -13.080 &        \\
$  2p^4(^3P)3s(^2P^e)$ & 28.180 &            &          &        \\
$  2p^4(^1D)3s(^2D^e)$ & 30.883 &    0.01469 &   -0.212 &   9.69 \\
$  2p^4(^3P)3p(^2D^o)$ & 31.416 &            &          &        \\
$  2p^4(^3P)3p(^2S^o)$ & 31.622 &            &          &        \\
$  2p^4(^3P)3p(^2P^o)$ & 31.745 &    0.02765 &  -13.143 &   5.71 \\
$  2p^4(^1S)3s(^2S^e)$ & 34.065 &    0.02787 &   -9.213 &        \\
$  2p^4(^1D)3p(^2F^o)$ & 34.261 &    0.00286 &   12.387 &        \\
$  2p^4(^1D)3p(^2P^o)$ & 34.416 &    0.10173 &   -8.303 &   0.78 \\
$  2p^4(^1D)3p(^2D^o)$ & 34.647 &            &          &        \\
$  2p^4(^3P)3d(^2D^e)$ & 34.928 &    0.00475 &   -7.571 &  12.80 \\
$  2p^4(^3P)3d(^2F^e)$ & 34.957 &            &          &        \\
$  2p^4(^3P)3d(^2P^e)$ & 35.053 &            &          &        \\
$  2p^4(^1S)3p(^2P^o)$ & 37.530 &    0.02213 &   -8.773 &   2.85 \\
$  2p^4(^1D)3d(^2G^e)$ & 37.983 &    0.00059 &  -17.893 &        \\
$  2p^4(^1D)3d(^2P^e)$ & 38.001 &            &          &        \\
$  2p^4(^1D)3d(^2S^e)$ & 38.038 &    0.00477 &  -21.942 &  10.14 \\
$  2p^4(^1D)3d(^2D^e)$ & 38.090 &    0.00701 &  -20.365 &   1.09 \\
$  2p^4(^1D)3d(^2F^e)$ & 38.152 &            &          &        \\
$  2p^4(^1S)3d(^2D^e)$ & 41.191 &    0.00817 &   -0.349 &        \\
$2s2p^5(^3P)3s(^2P^o)$ & 53.693 &    0.00567 &  -17.940 &        \\
$2s2p^5(^3P)3p(^2D^e)$ & 56.625 &    0.00145 &  -32.706 &        \\
$2s2p^5(^3P)3p(^2P^e)$ & 56.718 &            &          &        \\
$2s2p^5(^3P)3p(^2S^e)$ & 57.286 &    0.01115 &  -38.273 &        \\
$2s2p^5(^3P)3d(^2F^o)$ & 60.013 &    0.00024 &  -33.571 &        \\
$2s2p^5(^3P)3d(^2P^o)$ & 60.053 &    0.00105 &  -55.040 &        \\
$2s2p^5(^3P)3d(^2D^o)$ & 60.121 &            &          &        \\
$2s2p^5(^1P)3s(^2P^o)$ & 63.969 &    0.02148 &  -20.014 &        \\
$2s2p^5(^1P)3p(^2D^e)$ & 67.041 &    0.00028 &  -65.091 &        \\
$2s2p^5(^1P)3p(^2P^e)$ & 67.305 &            &          &        \\
$2s2p^5(^1P)3p(^2S^e)$ & 69.400 &    0.00173 &  -54.448 &   7.10 \\
$2s2p^5(^1P)3d(^2F^o)$ & 70.793 &    0.00030 &  -44.749 &        \\
$2s2p^5(^1P)3d(^2P^o)$ & 70.811 &    0.00067 &   -6.027 &  14.36 \\
$2s2p^5(^1P)3d(^2D^o)$ & 70.925 &            &          &        \\
$2s^02p^63s(^2S^e)$    & 87.663 &            &          &        \\
$2s^02p^63p(^2P^o)$    & 91.058 &            &          &        \\
$2s^02p^63d(^2D^e)$    & 94.579 &            &          &        \\
\end{tabular}
\end{ruledtabular}
\end{table}}

\section{Time delays for the $2s$ and $2p$ main lines}\label{sec:results}
\begin{figure}[tb]
  \includegraphics[width=\linewidth]{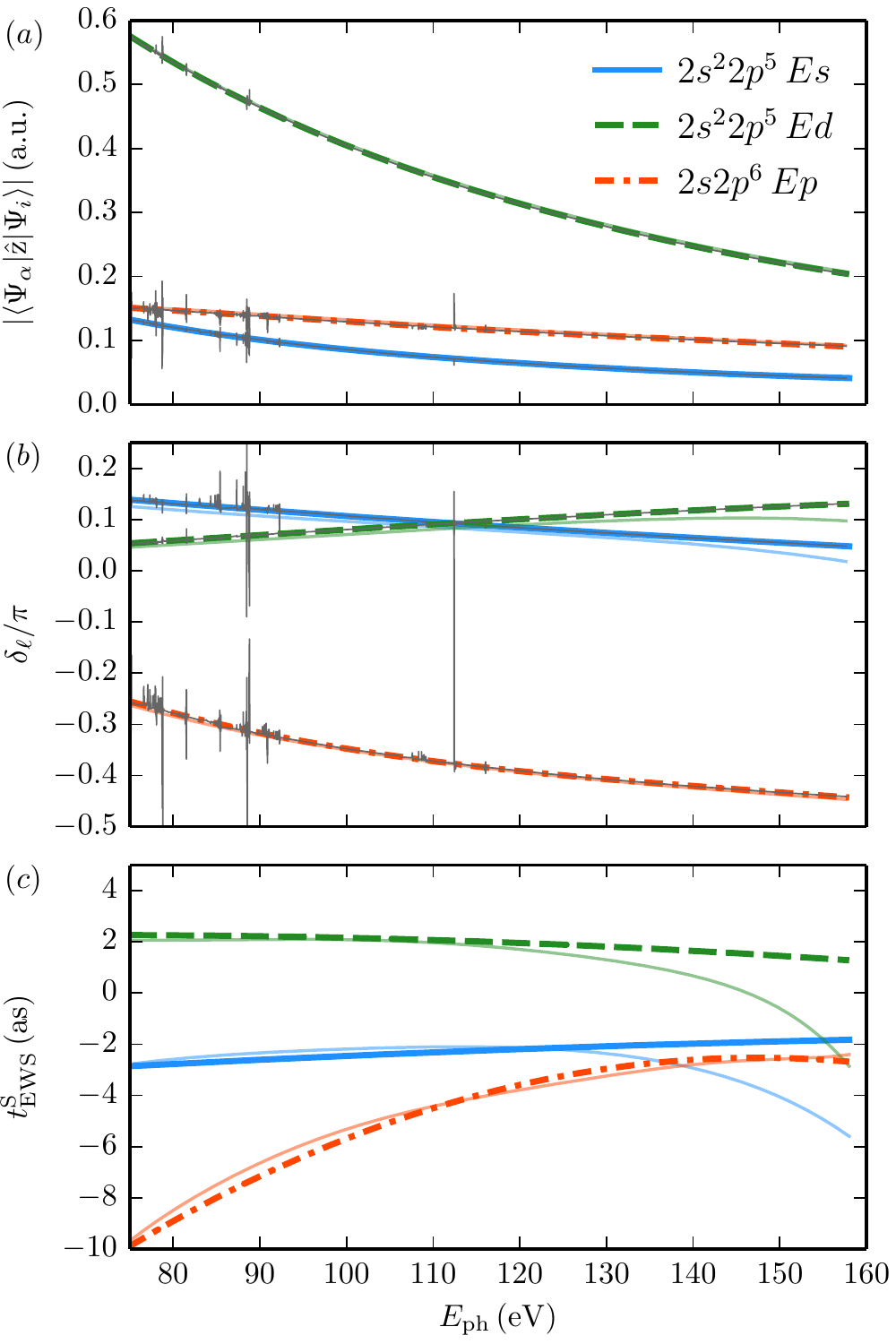}
  \caption{(Color online) (a) Modulus, (b) short-range phases in units of $\pi$, and (c) short-range contribution to the 
  EWS delay $\tEWSS$ for the dipole photo\-ionization matrix elements
    $\langle \Psi_\alpha | \op z | \Psi_i\rangle$ from the Ne ground state (\autoref{eq:matel_z}).
    For each channel, the darker colored line corresponds to the calculation with 38 target states after 
    removal of long-lived resonances (see text),
    while the lighter solid line corresponds to the 2-state calculation. The thin dark grey lines in (a) and (b) 
    show the ``raw'' 38-state data before removal of the resonances.}
  \label{fig:Ne_redmats}
\end{figure}
\begin{figure}[tb]
	\centering
		\includegraphics[width=\linewidth]{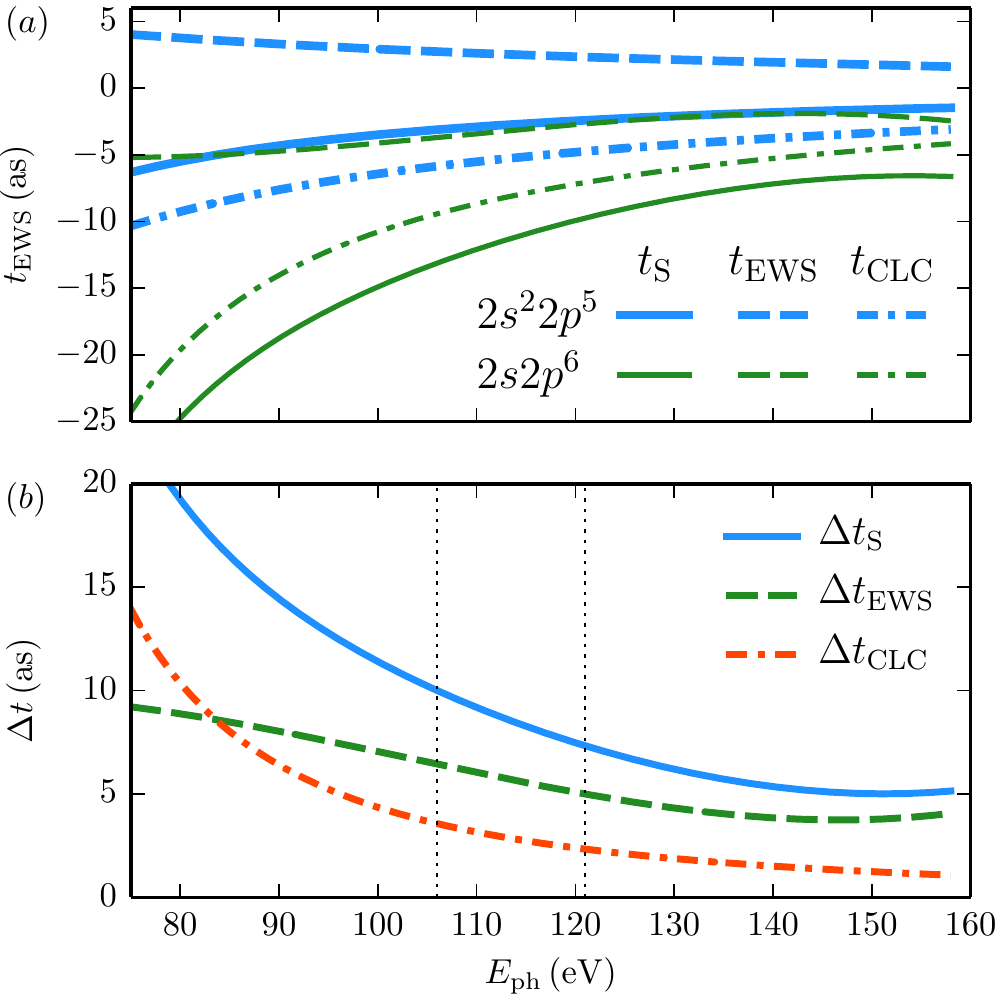}
	  \caption{(Color online) (a) Absolute time delays for photoionization of Ne with electron emission along the $z$ axis. 
            Solid lines: Full streaking time delay $\tS = \tEWS + \tCLC$. Dashed lines: $\tEWS$. Dash-dotted lines: $\tCLC$. 
            Blue: Ionization to the $2s^22p^5$ Ne$^+$ ionic state, green: ionization to the $2s2p^6$ state.
            (b) Relative streaking delay, $\Delta \tS=\Delta \tEWS + \Delta \tCLC$, between ionization to 
            $2s^22p^5$ and $2s2p^6$, and the relative contributions from CLC and EWS delays. Note that the 
            delays are given as a function of the photon energy; the kinetic energy of the different channels is therefore different.
          All results are obtained with 38 target states.}
\label{fig:delays}
\end{figure}

The moduli $\abs{\bra{\Psi_{\alpha}(E,\ell)}\op z\ket{\Psi_i}}$, short-range phases $\delta_\ell(E)$, 
and $\tEWSS(E)$ for the three exit channels contributing to the main lines, \ie without shake-up, 
for photoionization of the $2s$ and $2p$ electrons are displayed in \autoref{fig:Ne_redmats}. 
For photoionization of the $2p$ electron, two partial waves contribute with strong dominance of the $Ed$ 
channel (\autoref{fig:Ne_redmats}a). This is in line with the well-known $\ell\rightarrow \ell+1$ propensity rule~\cite{Fan1985}.
The short-range scattering phases  (\autoref{fig:Ne_redmats}b) vary, in the absence of resonances, 
only weakly over a wide range of photon energies ($80\eV\leq \Eph\leq160\eV$), resulting in a $\tEWSS$ 
contribution of typically less than $10\as$ (\autoref{fig:Ne_redmats}c).
The resulting total EWS delay and streaking delays (\autoref{fig:delays}a) for the $2s$ and $2p$ 
electrons vary somewhat stronger ($\leq 20\as$) over the same energy range. 
The major contribution comes from the CLC contribution (\autoref{fig:delays}a), which scales as a 
function of the kinetic energy of the outgoing electron $E_e$ as $\tCLC \sim -E_e^{3/2}$. 
Because of the difference in the ionization potentials and, consequently, in the kinetic energy of 
the outgoing electron, $\abs{\tCLC(2p)}$ is smaller than $\abs{\tCLC(2s)}$ at a given photon energy $\Eph$. 
The CLC contribution has been obtained from a highly accurate \emph{ab initio} time-dependent 
simulation of the streaking process for a hydrogen atom (\autoref{fig:delays}a), because it  
has its origin in the long-range, asymptotic $1/r$, \emph{hydrogenic}, Coulomb potential of the residual Ne$^+$ ion.
Alternatively, the CLC component could also be accurately determined by a purely classical trajectory 
analysis \cite{NagPazFei2011,PazNagBur2013}. 

The relative streaking time delay between the emission of $2s$ and $2p$ electrons, 
\begin{equation}\label{eq:delta_ts}
\Delta \tS = \tS^{\scriptstyle{(2p)}} - \tS^{\scriptstyle{(2s)}} \, ,
\end{equation}
was measured in the experiment \cite{SchFieKar2010} at photon energies of $106\eV$ ($\Delta\tS = 21 \pm 5\as$) and $121\eV$ ($\Delta\tS = 23 \pm 12\as$)
(vertical lines in \autoref{fig:delays}b). 
The theoretical calculations in \cite{SchFieKar2010} already showed that the electronic 
wavepacket emitted from the $2s$ shell precedes that of the $2p$ shell.
The present calculation yields (\autoref{fig:delays}b) $\Delta \tS=10.0\as$ at $106\eV$ and 
$\Delta \tS=7.3\as$ at $121\eV$, consisting of an EWS delay $\Delta \tEWS=6.4\as$ at $106\eV$ 
and $\Delta \tEWS=5.0\as$ at $121\eV$, respectively, and a CLC contribution of $\Delta \tCLC=3.6\as$ at $106\eV$ and $\Delta \tCLC=2.3\as$ at $121\eV$.
For $106\eV$, the EWS delay compares well with the $6.4\as$ obtained within the state-specific 
expansion approach~\cite{SchFieKar2010,MerKomNic2010}, but it is slightly lower than the $8.4\as$ obtained 
in a random-phase approximation with exchange~\cite{KheIva2010,Khe2013}.

The static dipole polarizabilities of the initial state Ne($2s^2 2p^6$), $\alpha\subscr{d}=2.65\au$ and
of the ionic final states $2s^2 2p^5$, $\alpha\subscr{d}=1.29\au$, and $2s2p^6$, $\alpha\subscr{d}=1.48\au$, which
are well reproduced by the present calculations, are far too small to lead to significant
quadratic Stark shifts even for strong streaking fields ($\approx 10\,$meV for $10^{13}\Wcm$).
Moreover, since the initial and final states are non-degenerate with sizable excitation gaps, 
the additional IR-field-induced contribution $\tdLC$ (\autoref{eq:delay_rel_all}) vanishes, i.e., $\tdLC=0$.

\begin{figure}[tb]
	\centering
		\includegraphics[width=\linewidth]{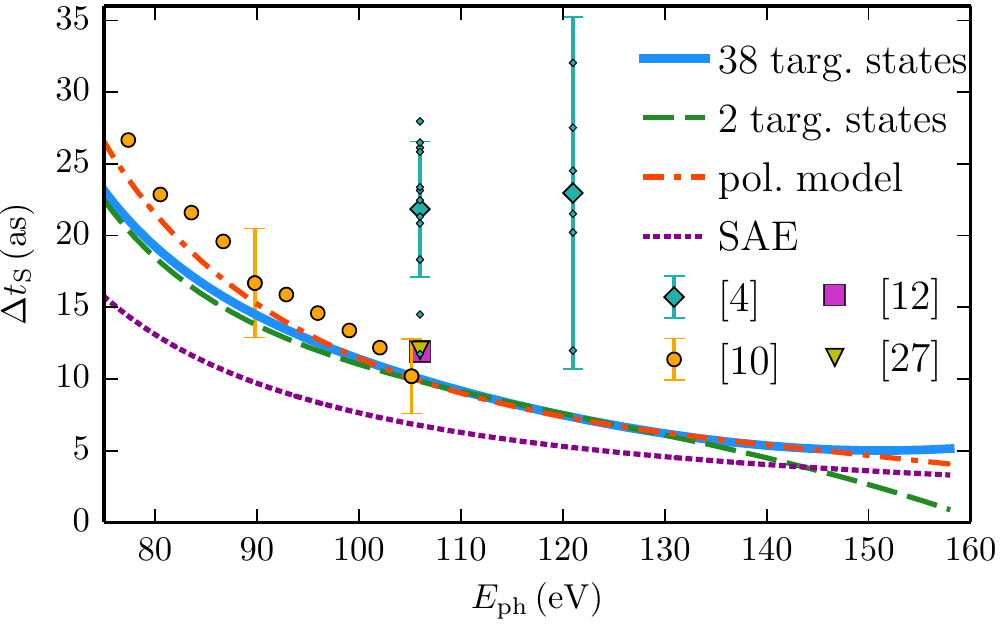}
	  \caption{(Color online) Comparison of results from three different basis sets used for the calculation of streaking delays. 
	  The model with two target states includes only the three channels of interest, while 
	  the calculation with 38 target states (57 channels) includes excited (shake-up) states in Ne$^+$ as 
	  well. The ``polarization model'' includes pseudo\-states to 
	  reproduce the polarizability of the ionic states. The predictions from all calculations agree very well 
	  within the energy range of interest and with the predictions of Moore \etal\cite{MooLysPar2011} 
	  within their error estimates (orange circles). They also agree reasonably well with the calculations
          of Kheifets~\cite{Khe2013} (magenta square) and Dahlstr\"om \etal\cite{DahCarLin2012} (yellow triangle),
	  while they are all much below the experimental values of Schultze \etal \cite{SchFieKar2010} (blue diamonds
          -- both the mean value and standard deviation and the individual data points as small dots on or near the error bars). The single-active-electron (SAE) results are
	  taken from~\cite{NagPazFei2011}. }
\label{fig:delays_comp}
\end{figure}

We find that the EWS delays and thus also the predicted streaking time shifts 
(see \autoref{eq:delay_rel_all}) are remarkably insensitive to the improvements of the basis
discussed above, especially in the experimentally relevant spectral region around $100\eV$ (\autoref{fig:delays_comp}).
Specifically for $\Delta\tS$ at $106\eV$, we obtain $9.82\as$, $10.00\as$, and $9.87\as$ from 
calculations with 2 target states, 38 target states, and 2 target states plus pseudo\-states (to 
account for polarizability effects), respectively.
The error of the extraction procedure, including the fitting of the phases to 4th-order polynomials, is approximately $\pm0.2\as$.
We thus conclude that our results for both the phases and the time delays are well-converged 
and that the electronic correlation in the ten-electron system is very well represented by the BSR method.
In \autoref{fig:delays_comp}, we compare our present calculations with the experimental data of 
Schultze \etal \cite{SchFieKar2010} as well as other theoretical results which include the influence of the IR field, specifically 
those by Moore \etal\cite{MooLysPar2011}, Kheifets and Ivanov~\cite{KheIva2010,Khe2013}, and Dahlstr\"om \etal\cite{DahCarLin2012}. 
Moore \etal employed the \hbox{$R$-matrix} incorporating time (RMT) approach with limited basis size, 
while the total delay in \cite{Khe2013} was obtained by adding $\Delta\tCLC$ from \cite{NagPazFei2012} 
to the EWS delay obtained using the random-phase approximation with exchange.
The delay in \cite{DahCarLin2012} was calculated using a diagrammatic technique for a two-photon matrix element 
relevant for the RABBIT technique, which gives equivalent results to attosecond streaking in smooth regions 
of the spectrum and also incorporates the CLC contribution (denoted $\tau\subscr{cc}$ in that context).
We also compare to a TDSE simulation in the single-active electron (SAE) approximation \cite{NagPazFei2012} 
for a Ne model potential where the electronic interactions are taken into account only at the mean-field level \cite{TonLin2005b}.
The present results are very close to the RMT prediction, in particular at the highest energy given 
in \cite{MooLysPar2011}, while the significant difference from the SAE model reflects the improved 
treatment of electronic correlation in the BSR approach. 
The close proximity to the RMT calculation underscores that full simulations of the streaking 
process are indeed not required if the additivity of EWS and IR-field-induced delays hold.
However, all theoretical results so far lie far off the experimental values by Schultze \etal \cite{SchFieKar2010} 
and are outside one standard deviation of all measured data points (\autoref{fig:delays_comp}). 
One should also note that all contributions to photoionization time delays decrease with increasing energy,
while no clear trend is recognizable in the experimental data. 
Summarizing the present analysis, the current state-of-the-art atomic theory of photoionization cannot 
fully account for the measured streaking delay between the $2s$ and $2p$ main lines of neon. 
A discrepancy of about $10\as$ (i.e., $50\%$ of the measured value) remains for photon energies $\geq 100\eV$.  

\section{Influence of shake-up channels}\label{sec:results_shake}
A possible source for the deviations could be the contamination of the streaking spectrum for 
the $2s$ main line in the experiment by unresolved shake-up channels. 
The latter can appear when the spectral width of the XUV pulse, $\Delta \omega\subscr{XUV}\approx2\pi/\tau\subscr{XUV}$, 
is larger than the spectral separation between the shake-up lines (``correlation satellites'') and the main line. 
For $\approx\!200\as$ XUV pulses, the width $\Delta\omega\subscr{XUV}$ is $\approx\!\!10\eV$.
We have recently shown that ionic shake-up channels can influence the extracted streaking delay significantly.
Specifically, in helium the streaking delay in the $n=2$ ionic channels is quite different from the streaking 
delay of all $n \geq 2$ channels together, even though the absolute yield is dominated by $n=2$ \cite{PazFeiNag2012}.

\begin{figure*}[tb]
\includegraphics[width=\linewidth]{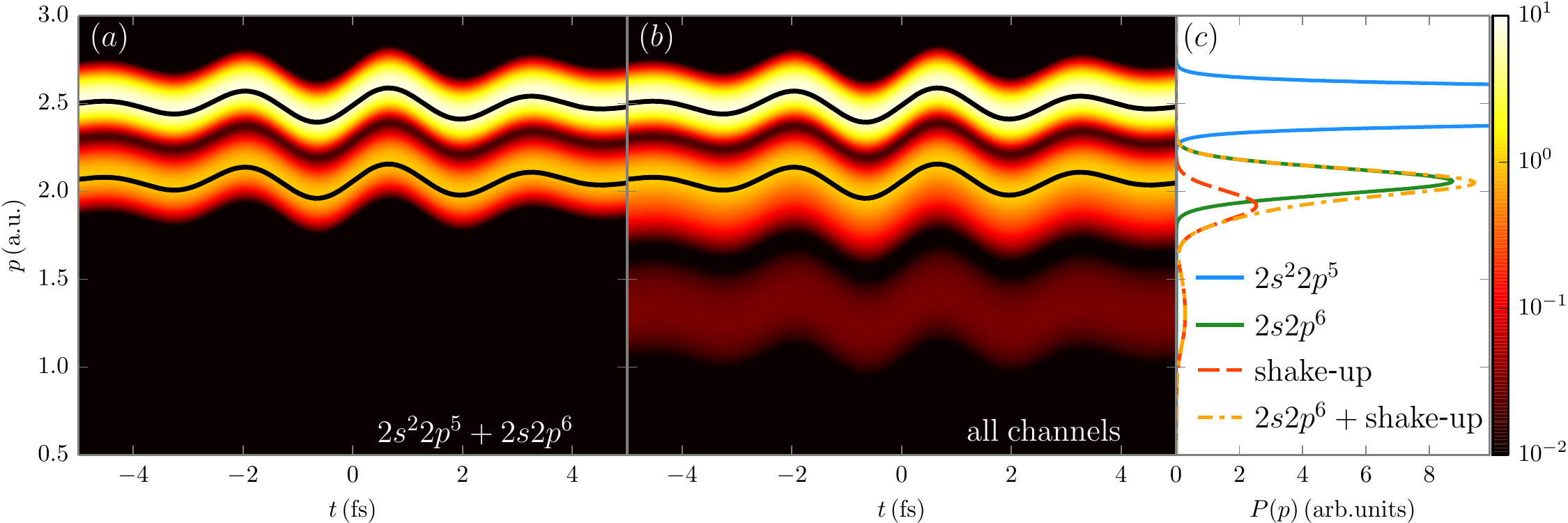}
\caption{(Color online) Influence of shake-up states on streaking spectra. All results are from the calculations 
with 38 target states (57 channels). (a) Simulated streaking spectrum according to \autoref{eq:synth_streaking} 
(see text for details) for the main lines including only electrons from the $2s^22p^5$ and $2s2p^6$ channels. 
(b) Simulated 
streaking spectrum including shake-up. (c) Unperturbed spectrum ($\vec{F}\subscr{IR}=0$) showing the 
contributions from the different channels. Shake-up denotes the sum over all shake-up channels.}
\label{fig:spectrogram}
\end{figure*}

The potentially strong influence of shake-up channels results from the prevalence of near-degenerate 
states in excited-state manifolds of the residual ion. 
Consequently, the ionic shake-up final state can be strongly polarized by the probing IR pulse 
\cite{BagMad2010,*BagMad2010Err,PazFeiNag2012,PazNagBur2013}, unlike for the ground state discussed above.
In this case the resulting dipole-laser coupling in the presence of the streaking field 
leads to a time-dependent energy shift due to a (quasi) linear Stark effect [proportional to 
the electric field strength $\vec{F}\subscr{IR}(t)$] and, in turn, an additional time shift. 
The exactly degenerate hydrogenic $\Hep$ residual ion is the prototypical case \cite{PazFeiNag2012}.
The quadratic Stark shift would lead to a time-dependent energy shift proportional 
to $F\subscr{IR}(t)^2$, which does not give rise to an additional IR-field-induced time delay. 
However, for near-degenerate states of opposite parity with an energy splitting $\Delta E\lesssim\omega\subscr{IR}$, 
the presence of a dLC contribution can be expected. 
To simulate and estimate the influence of shake-up lines on the neon 
spectrum (\autoref{fig:spectrogram}), we calculate the corresponding 
photoionization cross sections accompanied by shake-up and convolute them with a 
Gaussian frequency spectrum of an $106\eV$ XUV pulse with the experimental width (\autoref{fig:spectrogram}).
The sum over all shake-up channels (all states in \autoref{tab:states} except 
for the main lines) results in a sizable peak that significantly overlaps
with the $2s2p^6$ peak (corresponding to direct ionization of the $2s$ electron).
Such a contribution might significantly affect the experiment if it is not spectrally separated from the main line.
We note that in the experimental data (Fig.~2 of~\cite{SchFieKar2010}), a shoulder most likely due to shake-up is, indeed, visible.

In order to estimate the influence of shake-up on the streaking spectrogram for the $2s2p^6$ main line,
we synthesize a streaking spectrogram for a limited set of shake-up (SU) channels by including all 
excitation channels from the 38-state calculation that give a contribution along the $z$-direction (see \autoref{tab:states}).
The streaking scan is approximated as
\begin{equation}\label{eq:synth_streaking}
P_{\mathrm{S}}(t,p) = \sum_\alpha P_\alpha \; G(p,p_{\mathrm{S},\alpha}(t),\sigma_\alpha)
\end{equation}
with 
\begin{equation}\label{eq:streaking_formula}
p_{\mathrm{S},\alpha}(t) = p_{0,\alpha} + A(t+t_{\mathrm{S},\alpha}),
\end{equation}
where $p$ is the free electron momentum and $G(p,p_0,\sigma)$ is a normalized Gaussian centered at $p_0$ with standard deviation $\sigma$, while $A(t)$
is the vector potential of the streaking field. $P_\alpha\propto\abs{\bra{\Psi_\alpha}\op z\ket{\Psi_i}}^2$ is the 
ionization probability and $p_{0,\alpha} \!=\! \sqrt{2(\omega-E_\alpha)}$
is the momentum of the emitted electron at a photon energy of \hbox{$\hw=106\eV$}, where $E_\alpha$ 
is the ionization potential for 
reaching the final ionic state $\alpha$. The width $\sigma_\alpha = \sigma\subscr{XUV} / p_{0,\alpha}$ follows 
from assuming a constant spectral width $\sigma\subscr{XUV} = 4\eV$, corresponding to an XUV 
pulse with an intensity FWHM of $194\as$, close to the XUV pulse properties in the experiment \cite{SchFieKar2010}. 
Neglecting for the moment the presence of near-degenerate states in the ionic-state manifold 
accessed by shake-up (\autoref{tab:states}), the streaking shift $t_{\mathrm{S},\alpha}$ for 
each shake-up channel is given by \autoref{eq:delay_rel_all} with $\tdLC=0$.
Extracting the relative streaking time shifts from the resulting spectrogram for the 
synthesized streaking data shown in \autoref{fig:spectrogram}(b)
yields an estimated absolute delay for the resulting peak associated with $2s$ of 
$t_{\subscr{S,SU}} = -11.92\as$, compared to the delay of the $2s2p^6$ main line $\tS=-13.11\as$.
Accordingly, the effective $2p-2s$ delay is \emph{reduced} to $\Delta \tS=8.87\as$. 
Shake-up contributions can thus indeed influence the observed time delay.

Within the simple model outlined above, the inclusion of shake-up channels 
decreases rather than increases the delay and hence does not improve the agreement with the experiment. 
However, it should be noted that this result depends strongly on the model assumptions.
Specifically, we have implicitly assumed that one can neglect the coupling of closely 
spaced ionic shake-up states by the IR streaking field, and we have incoherently summed 
over the streaking contributions of individual shake-up states. 
Consequently, we go
another step further by taking into account the dynamical polarization due to closely spaced states. 
Two-state model calculations (not shown) have demonstrated that states with an energy 
difference much smaller than the IR photon energy (i.e., $\Delta E\ll\omega\subscr{IR}$) 
behave like degenerate states in an IR field. This results in an IR-field-induced dipole 
and additional timeshift $\tdLC$ (\autoref{eq:delay_rel_all}) for near-degenerate states. 
This behavior was confirmed in SAE streaking simulations with model potentials 
featuring near-degenerate states.
$\tdLC$ is determined by diagonalizing the dipole operator within a subspace of 
states with $|E_i - E_j| < \omega\subscr{IR}$ (for $\lambda=800\,$nm, i.e., $\hbar\omega\subscr{IR}\approx1.55\,$eV). 
In terms of the ``permanent'' (on the time scale of the IR field) dipole 
eigenstates $\Psi_k$ with dipole moment $d_k$, 
the amplitudes $\mu_{\alpha 0}$ of the matrix elements for the shake-up states $\Psi_\alpha$ 
with well-defined angular momentum can be written as
\begin{multline}\label{eq:exp_l_k}
\mu_{\alpha 0} = \bra{\Psi_{\alpha}(E,\Omega)} \op z \ket{0} = \\
\sum_k c_{\alpha k} \bra{\Psi_k(E,\Omega)} \op z \ket{0} = \sum_k c_{\alpha k} \mu_{k0}\, .
\end{multline}
Since the streaking setting with observation of ionization along the field axis 
breaks the rotational symmetry, states with well-defined angular momentum (in the absence of the 
streaking field) exhibit an effective dipole moment \cite{PazFeiNag2012}. This can be 
obtained by coherently summing the contributions from each dipole state $k$ with a 
Stark-like energy shift $d_k F\subscr{IR}(t)$ and is given by
\begin{equation}\label{eq:d_eff}
d_{\mathrm{eff},\alpha} = \Re\left( \frac{\sum_k d_k c_{\alpha k}\mu_{k0}}{ \mu_{\alpha 0} } \right)\,,
\end{equation}
resulting in a dipole-laser coupling induced time delay  
$\tdLC = \arctan(\omega\subscr{IR} d_{\mathrm{eff},\alpha} / p_0) / \omega\subscr{IR}$ \cite{BagMad2010,PazFeiNag2012,PazNagBur2013}. 

Taking these dipole-laser coupling contributions into account in the simulation of 
the streaking spectrogram leads to a positive contribution $\tdLC>0$ to the $2s$ delay 
and, hence, reduces its negative delay further to $t_{S,SU} = -11.27\as$ resulting 
in an effective relative $2p-2s$ delay of $\Delta \tS=8.22\as$. 
At this level of approximation, too, the discrepancy with experiment is (slightly) enhanced rather than reduced. 
For completeness, we add that for shake-up manifolds with resonant energy spacing 
($\Delta E\approx\omega\subscr{IR}$) of dipole-coupled states, single-active electron 
simulations indicate a further increase in $\tdLC$ due to coherent Rabi flopping 
dynamics by up to a factor of~5 compared to the degenerate case. 
Such a ``worst case scenario'', with 
$\tS \approx \tEWS + \tCLC + 5\tdLC$ for all states, would decrease the relative delay to $5.60\as$. 
Clearly, a more accurate determination requires a full quantum simulation of 
the streaking process for Ne shake-up channels.  This is presently out of reach.

\section{Summary and Conclusion}\label{sec:summary}
We have calculated streaking time shifts for the photoionization of $2s$ and $2p$ electrons in Ne, using 
highly accurate \hbox{$B$-spline} \hbox{$R$-matrix} models to obtain the Eisenbud-Wigner-Smith group delay of the electronic wavepackets 
and time-dependent streaking simulations to obtain the IR-induced contributions to the time shifts due to Coulomb and dipole-laser coupling.  
This method is expected to be superior to time-dependent methods that only take into account electronic interactions at the 
mean-field level and to time-independent calculations that neglect the influence of the infrared streaking field.
Since fully time-dependent calculations for many-electron systems are generally not yet feasible, such approaches 
are of pivotal importance for the understanding of time-resolved processes in complex systems.  
Our present results agree with predictions from other state-of-the-art calculations employing time-dependent $R$-matrix 
theory \cite{MooLysPar2011} for the relative $2p-2s$ time delay $\Delta \tS$ of the spectral main line.
The discrepancies with the experimental data remain. 
We identify unresolved contributions from the manifold of shake-up states as one possible source for the discrepancy. 
Our present estimates indicate, however, only moderate changes in $\Delta \tS$, which actually increase the discrepancy with the experimental data further.
Future experimental studies at different photon energies and for other atomic targets are therefore highly desirable.

\acknowledgments
This work was supported in part by the FWF-Austria (SFB NEXTLITE, SFB VICOM and P23359-N16), and by the 
United States National Science Foundation through a grant for the Institute for Theoretical Atomic, Molecular and Optical Physics at Harvard University and the Harvard-Smithsonian Center for Astrophysics, and through grants \hbox{No.~PHY-1068140} and \hbox{PHY-1212450} as well as XSEDE resources provided by NICS and TACC under Grant \hbox{No.~PHY-090031}. Some of the computational results presented here were generated on the Vienna Scientific Cluster (VSC). 
JF acknowledges support by the European Research Council under Grant \hbox{No.~290981} (PLASMONANOQUANTA).
RP acknowledges support by the TU Vienna Doctoral Program Functional Matter. 

\FloatBarrier
\bibliography{citeulike_atto_cleaned}

\end{document}